\documentclass[twocolumn,preprintnumbers]{revtex4}

\usepackage[dvipdfmx]{graphicx,color}

\usepackage{graphicx}
\usepackage{epsf}
\usepackage{amsmath,amssymb}

\usepackage{float}
\usepackage{color} 
\usepackage{multirow}
\usepackage{dcolumn}
\usepackage{array}
\usepackage{comment}

\usepackage{tikz}
\usepackage{array,booktabs}

\usepackage{ulem}

\def\vector#1{\mbox{\boldmath $#1$}}
\def\hana#1{\mathcal{#1}}
\def\Hesix{{}^6\textrm{He}}

\def\Ct{{}^{12}\textrm{C}}
\def\Cf{{}^{14}\textrm{C}}
\def\Cs{{}^{16}\textrm{C}}
\def\Os{{}^{16}\textrm{O}}
\def\Oe{{}^{18}\textrm{O}}
\def\Ot{{}^{20}\textrm{O}}
\def\2-b{\alpha + \Hesix}
\def\3-b{\alpha + \alpha + 2n}

\def\Kp{K^{\pi}}

\def\E1zero{1_{E1}^-(K=0)}

\usepackage[utf8]{inputenc}
\usepackage{natbib}

\begin{document}
\preprint{KUNS-2873}

\title{Low-energy dipole excitations in $^{20}$O with antisymmetrized molecular dynamics
}
\author{Yuki Shikata}
\author{Yoshiko Kanada-En'yo}
\affiliation{Department of Physics, Kyoto University, Kyoto 606-8502, Japan}


\begin{abstract}
Low-energy dipole (LED) excitations in $\Ot$ were investigated by 
variation after $K$-projection of deformation($\beta$)-constraint antisymmetrized molecular dynamics
combined with the generator coordinate method.
We obtained two LED states, namely, the $1_1^-$ state with 
one-proton excitation on the relatively weak deformation
and the $1_2^-$ state with a parity asymmetric structure of the normal deformation.
The former is characterized by a toroidal dipole (TD) mode with vortical nuclear current,
whereas the latter is associated with a low-energy $E1$ mode caused by surface neutron oscillation 
along the prolate deformation.
The TD (vortical) and $E1$ modes separately appear as the $\Kp=1^-$ and $\Kp=0^-$ components of  the deformed states, respectively, but couple with each other 
in the  $1_1^-$ and $1_2^-$ states of $\Ot$
because of $K$-mixing, and shape fluctuation. As a result of the mixing,  
 TD and $E1$ transition strengths are fragmented into the   $1_1^-$ and $1_2^-$ states.
We also obtained the $\Kp=0^+$, $\Kp=0^-$, and $\Kp=1^-$ bands with cluster structures
in the energy region higher than the LED states.
\end{abstract}

\maketitle
\section{Introduction}

Low-energy dipole (LED) excitations, that appear in an energy region lower than 
giant dipole resonances is a topic  
gaining attention in experimental and theoretical research
since a few decades
(see, reviews in 
Refs.~\cite{Paar:2007bk,1402-4896-2013-T152-014012,Savran:2013bha,Bracco:2015hca} and references).
Significant dipole strengths of several percent of the energy-weighted sum rule (EWSR) 
have been observed in stable nuclei in a wide mass-number range from  $^{12}$C to $^{208}$Pb
\cite{Harakeh:1981zz,Decowski:1981pcz,Poelhekken:1992gvp,John:2003ke}. 
LEDs were also discovered recently in neutron-rich nuclei such as $^{20}$O~\cite{Tryggestad:2002mxt,Tryggestad:2003gz,Nakatsuka:2017dhs}, $^{26}$Ne~\cite{Gibelin:2008zz}, and $^{48}$Ca~\cite{PhysRevLett.85.274,Derya:2014yqk}. 
To describe the isospin properties of the LED strengths,  
various types of dipole modes were theoretically proposed. For example, the neutron skin mode (Pigmy mode) \cite{Brzosko:pigmy,Ikeda:pygmy} was considered for 
$E1$ strengths  \cite{Paar:2007bk,1402-4896-2013-T152-014012,Savran:2013bha,Bracco:2015hca}.
To understand isoscalar LED strengths, toroidal (also called vortical or torus) were suggested\cite{Semenko:toroidal,Ravenhall:1987thb,Ryezayeva:2002zz,Paar:2007bk,Papakonstantinou:2010ja,Kvasil:2011yk,Repko:2012rj,Kvasil:2013yca,Nesterenko:2016qiw,Nesterenko:2017rcc}, and cluster modes were 
investigated~\cite{Chiba:2015khu,Kanada-Enyo:2017uzz,Kanada-Enyo:2017fps,Kanada-Enyo:2019hrm,Shikata:2019wdx,Shikata:2020lgo,Shikata:2020iez}.
However, the observed LED strengths were not fully described, and their origins are not clarified yet. 

LED excitations in oxygen isotopes have been intensively investigated in experimental and theoretical studies.
In theoretical studies based on mean-field approaches , 
LED strengths  were described as noncollective single-particle excitations 
on the spherical or slightly deformed ground state~\cite{Sagawa:1999zz,Colo:2001fz,Sagawa:2001jhf,Vretenar:2001hs,Paar:2002gz,Inakura:2018ccl}. 
However, cluster structures of neutron-rich O isotopes such as $\Oe$ and $\Ot$ have been discussed to describe 
excited states in the low-energy region~\cite{Gai:1983zz,Gai:1987zz,PhysRevC.43.2127,Furutachi:2007vz,Baba:2019csd,Baba:2020iln,Shikata:2020iez}.
On the experimental side, the isovector dipole strengths  for $^{17-22}$O were observed 
in excitation energy $E_x\lesssim 15$ MeV region, and were found to  
exhaust a few percentages of the Thomas–Reiche–Kuhn sum rule~\cite{Leistenschneider:2001zz}. 
In the case of $^{20}$O, the $E1$ and isoscalar dipole (ISD) transition strengths for the $1^-_1$  and $1^-_2$ states 
were measured~\cite{Tryggestad:2002mxt,Tryggestad:2003gz,Nakatsuka:2017dhs}
and indicated a difference of isospin properties between 
these two states.

In our previous study~\cite{Shikata:2020iez}, we investigated LED excitations in $\Oe$ using
variation after $K$-projection (K-VAP) in the framework of $\beta$-constraint 
antisymmetrized molecular dynamics (AMD)~\cite{KanadaEnyo:1995tb,10.1143/PTP.106.1153,Kanada-Enyo:2001yji,Kanada-Enyo:2012yif,Kimura:2016fce}
combined with the generator coordinate method (GCM), 
which was developed for the study of LED excitations in Ref.~\cite{Shikata:2020lgo}.
Two LED excitations were obtained in  $\Oe$. One is dominantly described with a one-particle one-hole (1p-1h) excitation on a weak deformation and
the other contains $\Cf+\alpha$ cluster component.
The $1_1^-$ and $1_2^-$ states had significant toroidal dipole (TD) and compressive dipole (CD) strengths through $K$-mixing and shape fluctuation along the deformation $\beta$.

In this study, we apply the same method of  K-VAP and GCM  of $\beta$-constraint AMD to $\Ot$. 
We focus on LED excitations and cluster states in the $\Ot$ system in particular.
The isospin properties of dipole transition strengths are investigated in detail, and the roles of excess neutrons in LED excitations of $\Ot$ are discussed with $\Os$ and $\Oe$.  

This paper is organized as follows.
In Sect.~\ref{sec:formalism}, the framework of K-VAP and GCM of $\beta$-constraint AMD is explained.
The definition of the dipole operators is also given. 
Section~\ref{sec:effective_int} describes the effective nuclear interaction used in the present calculation.
The calculated results for $\Ot$ are shown in Sect.~\ref{sec:results}.
In Sect.~\ref{sec:discussion}, the properties of LED excitations in $\Ot$ are analyzed in detail, and systematics of 
LED excitations along the isotope chain, $\Os$, $\Oe$, and $\Ot$ are discussed.
Finally, a summary is given in Sect.~\ref{sec:summary}.
In appendix \ref{app:1}, definitions of operators and matrix elements for densities and transitions 
are given.

\section{Formalism}\label{sec:formalism}

To investigate LED excitations, we apply K-VAP and GCM in the framework of $\beta$-constraint AMD to $\Ot$, just as we 
did in our previous work for $\Oe$~\cite{Shikata:2020iez}. For detailed formulation, the reader is referred to Ref.~\cite{Shikata:2020iez} and references therein.

We first perform energy variation for the $\beta$-constraint AMD wave function after $K$ and parity ($K^\pi$) projection. 
Following the K-VAP, we superpose the obtained basis wave functions with total-angular-momentum and parity ($J^\pi$) projection by solving 
the GCM (Hill-Wheeler) equation for $K$ and $\beta$, and obtain final results of the total wave functions and energy spectra of the $J^\pi_k$ states of $\Ot$.
The transition strengths are calculated for the $J^\pi_k$ states. 
For a detailed discussion, 
we also analyze each basis wave function in the intrinsic frame before the superposition.

An AMD wave function for $A$-body system $\Phi$ is expressed using a Slater determinant of single-particle wave functions~\cite{Kanada-Enyo:1998onp,Kanada-Enyo:2001yji}:
\begin{eqnarray}
\Phi = \hana{A}\left[\psi_1\psi_2\cdots\psi_A\right],
\end{eqnarray}
where $\psi_i$ represents the $i$th single-particle wave function written by a product of spatial, 
spin, and isospin functions as follows:
\begin{eqnarray}
\psi_i\ &=& \phi(\vector{Z}_i)\chi(\vector{\xi}_i)\tau_i, \\
\phi(\vector{Z}_i) &=& \left(\frac{2\nu}{\pi}\right)^{\frac{3}{4}}\exp\left[ -\nu\left(\vector{r} - \frac{\vector{Z}_i}{\sqrt{\nu}}\right)^2 \right],\\
\chi(\vector{\xi}_i)&=&\xi_{i\uparrow}|\uparrow\rangle + \xi_{i\downarrow}|\downarrow\rangle, \\
\tau_i &=&p\ \textrm{or}\ n.
\end{eqnarray}
Here, $\vector{Z_i}$ and $\vector{\xi_i}$ are complex variational parameters for Gaussian centroids and spin directions, respectively.
For the width parameter $\nu$, a fixed value is used for all nucleons.

In the K-VAP method, the energy variation is done for the parity- and $K$-projected AMD wave function $|\Psi\rangle = \hat{P}_K\hat{P}^{\pi}|\Phi\rangle$
with the quadrupole deformation $\beta$ constrains.
Here, $\hat{P}^{\pi}$  and $\hat{P}_{K}$  are the parity- and $K$-projection operators, respectively. 
In the present calculation, $\Kp=0^+,\ \Kp=0^-$, and $\Kp=1^-$ are adopted to 
obtain the $\Kp$ bases optimized for the ground and dipole states. 
The $\beta$ constraint is imposed for the AMD wave function
during the energy variation.
We follow the definition of the quadrupole deformation parameters 
$\beta$ and $\gamma$ adopted for the $\beta\gamma$-constraint AMD in Ref.~\cite{Suhara:2009jb}, however,
the constraint is imposed only on  $\beta$ but not on $\gamma$ in the present $\beta$-constraint AMD. 
It means that 
$\gamma$ is optimized for each $\beta$ and can be finite.

After K-VAP with each $\beta$ value, we obtained the optimized AMD wave functions 
$|\Phi_K^\pi(\beta)\rangle$ for $\Kp=0^+,\ \Kp=0^-$, and $\Kp=1^-$, which we call
as $K0^+(\beta)$, $K0^-(\beta)$, and $K1^-(\beta)$ bases, respectively. 
To obtain the total wave functions and energy spectra of the $J^{\pi}_k$ states of $\Ot$, the GCM calculation 
 is performed by superposing the basis wave functions along $\beta$ as
\begin{eqnarray}
|\Psi^{\pi}(J_k)\rangle = \sum_{K,K'}\sum_{\beta}c_{KK'}(\beta)\hat{P}_{MK'}^J\hat{P}^{\pi}|\Phi_K^{\pi}(\beta)\rangle ,
\end{eqnarray}
where $\hat{P}_{MK}^J$ is the angular-momentum-projection operator, and 
coefficients $c_{KK'}(\beta)$ are determined by diagonalizing the Hamiltonian and norm matrices. 
For the negative-parity states, $K$-mixing is taken into account and mixing of the $K0^-(\beta)$ and $K1^-(\beta)$ bases were
considered.

The total wave functions for the 
$J^{\pi}_k$ states obtained after GCM are used to calculate the  transition strengths such as the $E1$, $E2$, and $E3$ strengths.
For dipole transitions from the ground state, we consider three types of dipole operators, $E1$, TD, and CD operators
used in Refs.~\cite{0954-3899-29-4-312,Kvasil:2011yk}, 
\begin{eqnarray}
\hat{M}_{E1}(\mu) &=& \frac{N}{A}\sum_{i\in p}r_iY_{1\mu}(\hat{\vector{r}}_{i}) - \frac{Z}{A}\sum_{i\in n}r_iY_{1\mu}(\hat{\vector{r}}_{i}), \\
\hat{M}_{\textrm{TD}}(\mu)&=&\frac{-1}{10\sqrt{2}c} \int d\vector{r}\ (\nabla\times\hat{\vector{j}}_{\textrm{nucl}}(\vector{r}))\cdot r^3\vector{Y}_{11\mu}(\hat{\vector{r}}), \label{eq:TD_op} \nonumber \\ \\
\hat{M}_{\textrm{CD}}(\mu) &=& \frac{-1}{10\sqrt{2}c}\int d\vector{r}\ \nabla\cdot\hat{\vector{j}}_{\textrm{nucl}}(\vector{r})\ r^3Y_{1\mu}(\hat{\vector{r}}), \label{eq:CD_op}
\end{eqnarray}
where $\vector{Y}_{jL\mu}(\hat{\vector{r}})$ are vector spherical harmonics and 
$\hat{\vector{j}}_{\textrm{nucl}}(\vector{r})$ is the convection nuclear current defined by
\begin{eqnarray}
\hat{\vector{j}}_{\textrm{nucl}}(\vector{r}) &=& \frac{-i\hbar}{2m}\sum_{k=1}^A\left\{ \vector{\nabla}_k\delta(\vector{r}-\vector{r}_k) + \delta(\vector{r}-\vector{r}_k)\vector{\nabla}_k \right\}.  \label{eq:current}\nonumber \\
\end{eqnarray}
The $E1$ operator is the isovector dipole operator, whereas 
the TD and CD operators are isoscalar type operators that can measure the nuclear vortical and 
compressional dipole modes, respectively. 
For a dipole operator $D=\{E1, \textrm{TD},\textrm{CD}\}$, the transition strength $B(D;0_1^+\rightarrow 1_k^-)$ 
is given as $|\langle 1_k^-||\hat{M}_D||0_1^+\rangle|^2$.
Notably, the CD transition strength corresponds to the standard ISD transition strength
with the relation,
\begin{eqnarray}
B(\textrm{CD};0_1^+\rightarrow 1_k^-) &=& \left(\frac{1}{10}\frac{E_k}{\hbar c}\right)^2B(\textrm{ISD};0_1^+\rightarrow 1_k^-), \nonumber \\
\end{eqnarray}
where $E_k$ is the excitation energy of the $1_k^-$ state.

\section{Effective interaction} \label{sec:effective_int}
The effective Hamiltonian used in the present study is given as
\begin{eqnarray}
H = \sum_i t_i - T_G + \sum_{i<j}v_{ij}^{\textrm{coulomb}} + V_{\textrm{eff}}.
\end{eqnarray} 
Here, $t_i$ and $T_G$ are the kinetic energy of the $i$th nucleon and that of the center of mass, respectively,
 and $v_{ij}^{\textrm{coulomb}}$ is the Coulomb potential.
The effective nuclear potential $V_{\textrm{eff}}$ includes the central and spin-orbit potentials.
We use the MV1 (case 1) central force~\cite{Ando:1980hp} with the parameters $W = 1-M = 0.38$ and $B=H=0$, and 
the spin-orbit part of the G3RS force~\cite{Tamagaki:1968zz, Yamaguchi:1979hf} with the strengths $u_1=-u_2=-3000$\ MeV.
This set of parametrization is identical to that used for the AMD calculations of $p$-shell and $sd$-shell nuclei
in Refs.~\cite{Kanada-Enyo:1999bsw,Kanada-Enyo:2006rjf,Kanada-Enyo:2017ers,Kanada-Enyo:2020goh}.
It describes the energy spectra of $\Ct$ including $1^-$ states.
The width parameter is chosen as $\nu=0.16$ fm$^{-2}$, which reproduces the nuclear size of $^{16}$O 
with a closed $p$-shell configuration in the harmonic oscillator shell model.

\section{Results of $\Ot$} \label{sec:results}

\subsection{Energies and band structure}
\begin{figure}
\begin{center}
\includegraphics[width=\hsize]{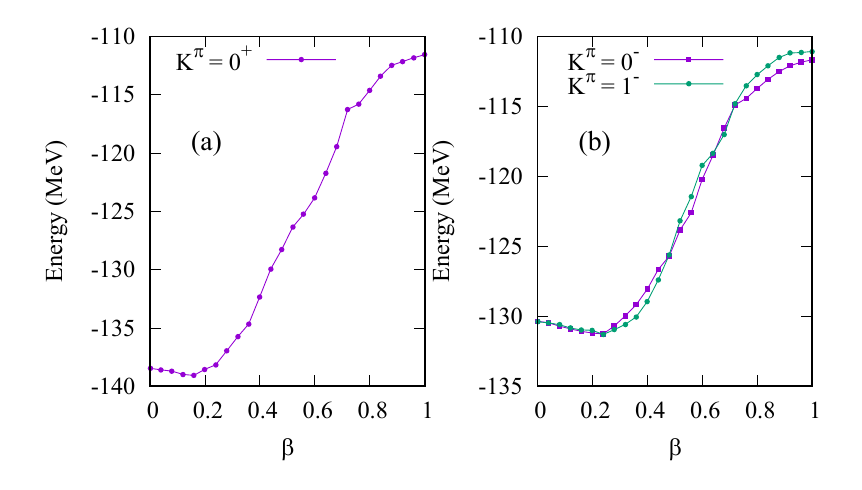}
\end{center}
\caption{
$K$-projected energy curves of $\Ot$ obtained by K-VAP of $\beta$-constraint AMD 
plotted as a function of quadrupole deformation $\beta$.
The energy curve for the $K0^+(\beta)$ bases
is shown in panel (a) and  those for the $K0^-(\beta)$ and $K1^-(\beta)$ bases
are shown in panel (b).
}
\label{fig:energy_surface}
\end{figure}

\begin{figure}
\begin{center}
\includegraphics[width=\hsize]{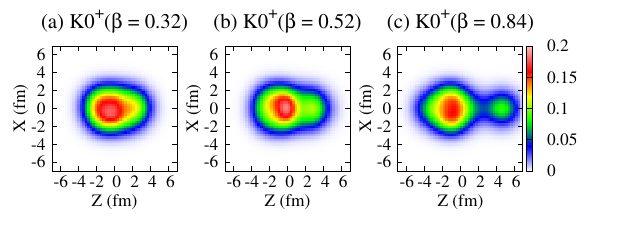}
\end{center}
\caption{
The matter densities of $K0^+$ bases at (a)~$\beta=0.32$, (b)~0.52, and (c)~0.84.
The intrinsic densities are integrated along the $Y$-axis and plotted on the $Z$-$X$ plane. 
The unit is fm$^{-2}$.
}
\label{fig:matter_density_positive}
\end{figure}

\begin{figure}
\begin{center}
\includegraphics[width=\hsize]{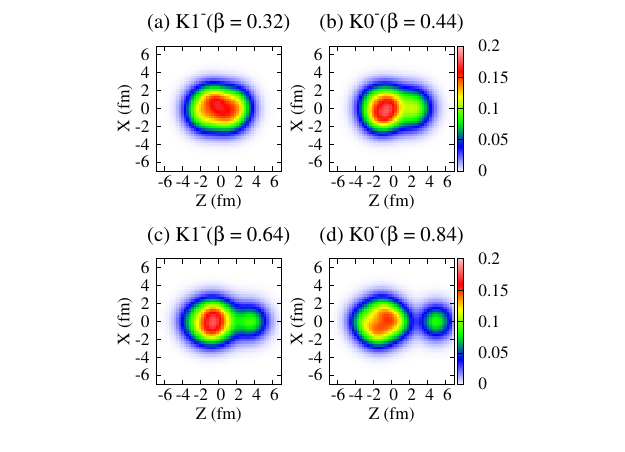}
\end{center}
\caption{
The same as Fig.~\ref{fig:matter_density_positive} but for the (a)~$K1^-(0.32)$,  (b)~$K0^-(0.44)$, 
 (c)~$K1^-(0.64)$, and (d)~$K0^-(0.84)$ bases.}
\label{fig:matter_density_negative}
\end{figure}

By performing the energy variation for the $K^\pi$-projected AMD wave function,
we obtain the wave functions of the $K0^+$, $K0^-$, and $K1^-$ bases at
each $\beta$ value. 
The $K^\pi$-projected energy curve of the $K0^+$ bases is shown in Fig.~\ref{fig:energy_surface}(a), 
whereas the $K0^-$ and $K1^-$ bases are shown in Fig.~\ref{fig:energy_surface}(b).
Energy minimums exist
around $\beta=0.2$ which corresponds to the intrinsic states of 
the ground $0^+$ and lowest $1^-$ states. In the larger $\beta$ region, there is no 
local minimum. The intrinsic structure changes with an increase of $\beta$ along the energy curves
as shown in Figs.~\ref{fig:matter_density_positive} and \ref{fig:matter_density_negative}, which 
display intrinsic matter density of typical positive- and negative-parity bases, respectively.
The intrinsic structure around  
the energy minimum has a weak deformation and changes 
to the prolate deformation with a cluster structure 
at $\beta=0.52$, and finally a developed cluster structure of $\Cs+\alpha$ appears in the $K0^+(0.84)$ base
as shown in Figs.~\ref{fig:matter_density_positive}(a), (b), and (c) 
for the $K0^+(0.32)$, $K0^+(0.52)$, and $K0^+(0.84)$ bases.
In the negative-parity case,
the $K0^-$ and $K1^-$ bases degenerate
in $\beta\lesssim$ 0.7 region (see Fig.~\ref{fig:energy_surface}(b)). 
Cluster structures appear in the 
$K0^-$ and $K1^-$ bases as $\beta$ increases. Because of this clustering, 
the $K0^-$ energy becomes lower than the $K1^-$ energy in the $\beta\gtrsim$ 0.7 region
because the developed cluster structure favors 
the $K0^-$ component.
In the large $\beta$ region, the negative-parity bases have the $\Cs+\alpha$ cluster structure similar
to the $K0^+$ bases (see Fig.~\ref{fig:matter_density_positive}(c) and  Fig.~\ref{fig:matter_density_negative}(d)).

\begin{figure}
\begin{center}
\includegraphics[width=\hsize]{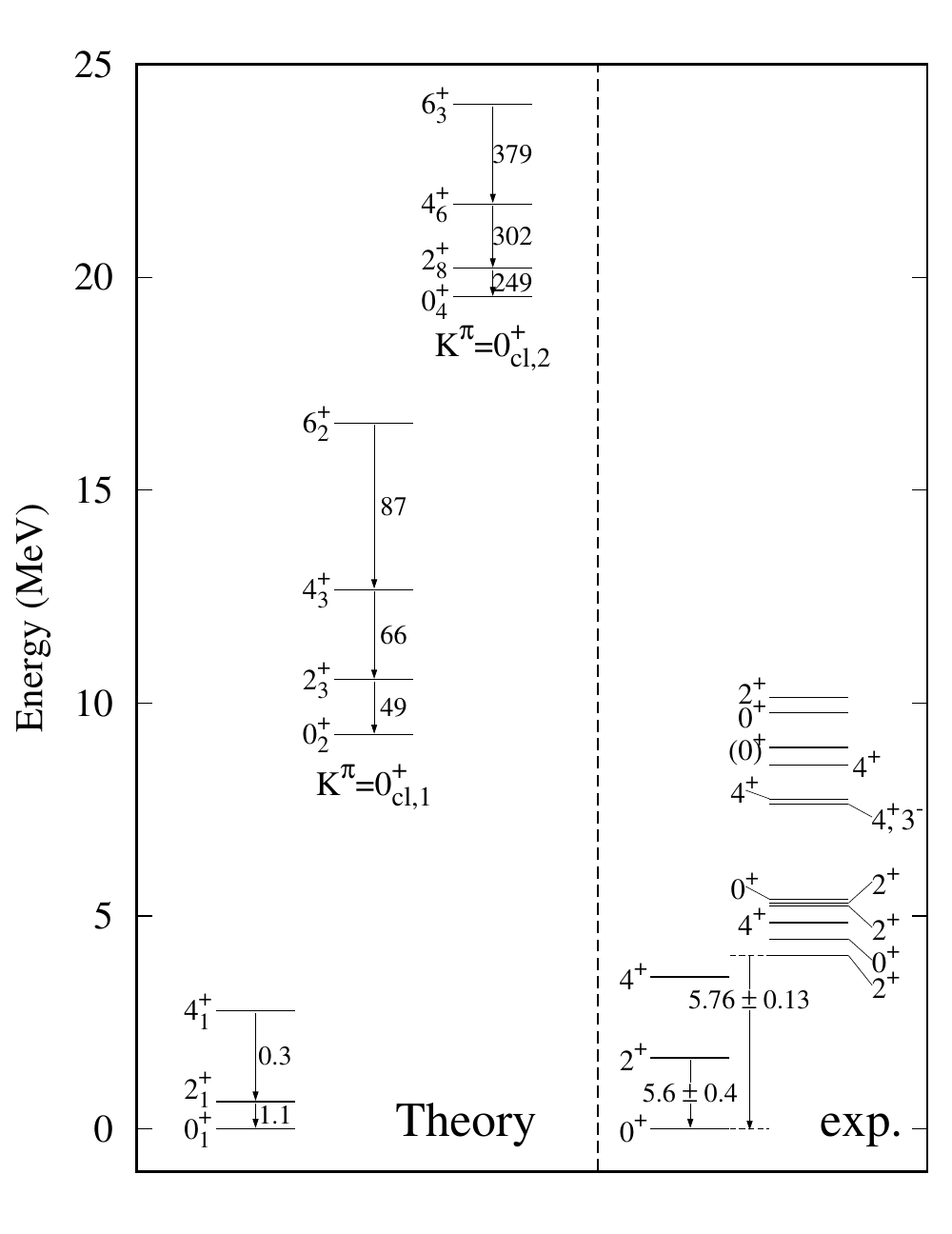} 
\end{center}
\caption{
Energy spectra of positive-parity states in $\Ot$ obtained by GCM
and those of experimental data.
In the calculated spectra, the $\Kp=0^+_1$, $\Kp=0_\textrm{cl,1}^+$, and $\Kp=0_\textrm{cl,2}^+$ bands 
are shown together with the $B(E2)$ values of in-band transitions.
The experimental $B(E2)$ values are taken from Refs.~\cite{Nakatsuka:2017dhs,Tryggestad:2003gz}.
The unit of $B(E2)$ is $e^2$fm$^4$.
}
\label{fig:energy_spectrum_positive}
\end{figure}

\begin{figure}
\begin{center}
\includegraphics[width=\hsize]{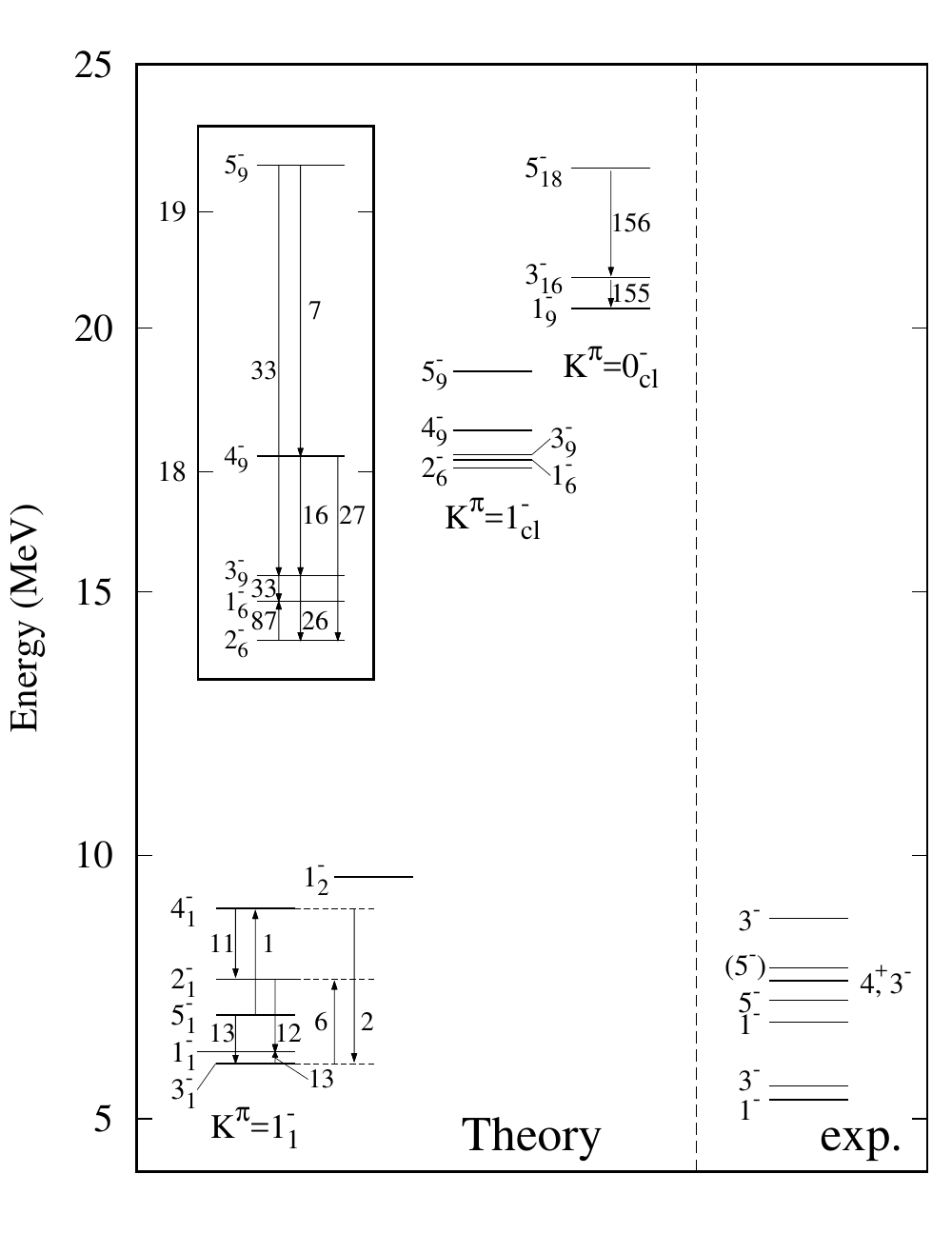} 
\end{center}
\caption{
Energy spectra of negative-parity states in $\Ot$ obtained by GCM
and experimental negative-parity spectra.
In the calculated result, spectra of the $\Kp=1^-_1$, $\Kp=1_\textrm{cl}^+$, and $\Kp=0_\textrm{cl}^+$ bands
and that of the $1^-_2$ state are shown together with the $B(E2)$ ($e^2$fm$^4$) values of in-band transitions.
For the $\Kp=1_\textrm{cl}^+$ band, spectra on a  large scale are inserted in the figure. 
}
\label{fig:energy_spectrum_negative}
\end{figure}

\begin{figure}
\begin{center}
\includegraphics[width=\hsize]{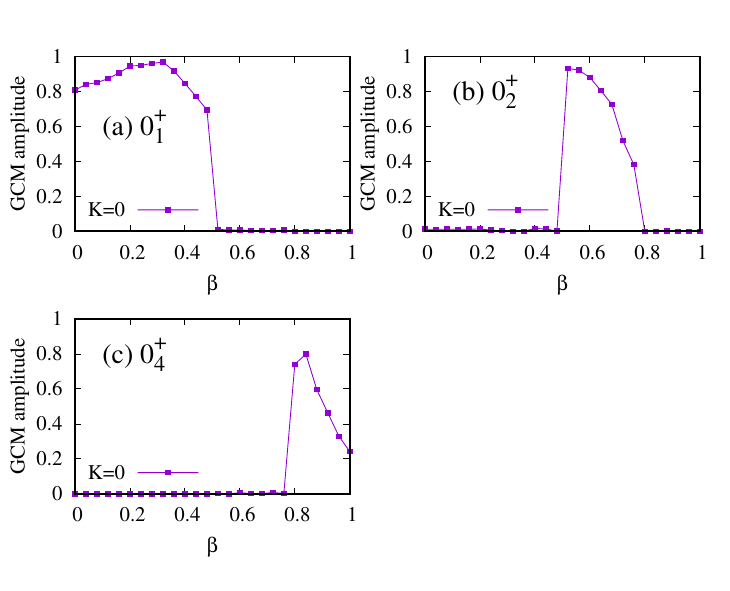} 
\end{center}
\caption{
GCM amplitudes of the positive-parity states.
The amplitudes calculated by 
squared overlap with the $K0^+(\beta)$ bases are plotted as a function of $\beta$.
The results for the band-head $0_1^+$, $0_2^+$, and $0_4^+$ states of the 
$\Kp=0^+_1$, $\Kp=0^+_\textrm{cl,1}$, $\Kp=0^+_\textrm{cl,2}$ bands are shown in panels 
(a), (b), and (c), respectively. 
}
\label{fig:GCMamplitude_positive}
\end{figure}

\begin{figure}
\begin{center}
\includegraphics[width=\hsize]{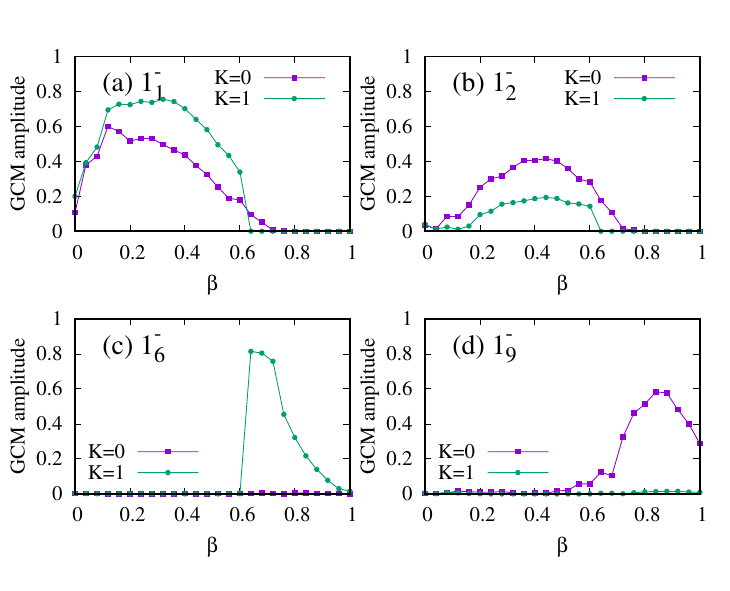} 
\end{center}
\caption{
GCM amplitudes of the negative-parity states.
The amplitudes calculated by 
squared overlap with the $K0^-(\beta)$ and $K1^-(\beta)$ bases are plotted by
squares and circles, respectively. 
Panels (a), (c), and (d) show the results for
the band-head $1_1^-$, $1_6^-$, and $1^-_9$ states of the 
$\Kp=1^-_1$,  $\Kp=1^-_\textrm{cl}$, and  $\Kp=0^-_\textrm{cl}$ bands, respectively,
and panel (b) shows the result for the $1_2^-$ state. 
}
\label{fig:GCMamplitude_negative}
\end{figure}

The energy spectra of $\Ot$ are obtained
after the GCM calculation using the basis wave functions obtained by K-VAP of $\beta$-constraint AMD.
The calculated binding energy is $141.7$~MeV, which is slightly smaller than the experimental value
($151.36$~MeV). 
The positive- and negative-parity energy spectra are shown in Fig.~\ref{fig:energy_spectrum_positive}
and Fig.~\ref{fig:energy_spectrum_negative}, respectively. To discuss band structure, 
we show theoretical energy spectra for band member states, which can be classified into $K^\pi=0^+$, $K^\pi=0^-$, and
$K^\pi=1^-$ bands, and that for the $1^-_2$ state 
along with the calculated $B(E2)$ values of in-band transitions on the left, and  
the experimental energy spectra and $B(E2)$ values on the right of the figures.
Figures \ref{fig:GCMamplitude_positive} and \ref{fig:GCMamplitude_negative} show the GCM amplitudes
for the band-head states, which are
defined by squared overlap with each base.

The ground band ($K^\pi=0^+_1$ band), which consists of the $0_1^+$, $2_1^+$, and  $4_1^+$ states 
is constructed from the basis wave functions with weak deformation around the energy minimum
of the $K^\pi=0^+$ energy curve. The calculated $E2$ transition strengths in the ground band 
are small because of a proton shell closure feature. This result is qualitatively consistent with 
the experimental $B(E2)$ value, but quantitatively, underestimates 
the observed data.

Above the ground band, two $K^\pi=0^+$ bands are built on the 
$0^+_2$ and $0^+_4$ states showing cluster structures. The lower and higher cluster
bands on the $0^+_2$ and $0^+_4$ states are labeled as 
$K^\pi=0^+_{\textrm{cl},1}$ and $K^\pi=0^+_{\textrm{cl},2}$ bands, respectively.  
The former ($K^\pi=0^+_{\textrm{cl},1}$) band is mainly formed by the 
$K0^+(0.52)$ base (Fig.~\ref{fig:matter_density_positive}(b)), which has a deformed structure with clustering.
The latter ($K^\pi=0^+_{\textrm{cl},2}$) band contains the dominant component of the 
$K0^+(0.84)$ base~(Fig.~\ref{fig:matter_density_positive}(c)) with a developed $\Cs+\alpha$ cluster structure.
Because of the largely deformed intrinsic structure for 
these cluster bands, strong $E2$ transitions are obtained for in-band transitions, in particular, 
in the $K^\pi=0^+_{\textrm{cl},2}$  band.

In the calculated negative-parity levels, 
we obtain the $1^-_1$ and $1^-_2$ states in the low-energy region (see Fig.~\ref{fig:energy_spectrum_negative}).
The $K^\pi=1^-_1$ band is built on the $1^-_1$ state, whereas  the 
$1^-_2$ state does not form a clear band structure. 
As shown in Fig.~\ref{fig:GCMamplitude_negative}, the
$1^-_1$ and $1^-_2$ states in the low-energy region 
contain components of basis wave functions in the $\beta\lesssim 0.6$ regions corresponding 
to weak or normal deformations shown in Figs.~\ref{fig:matter_density_negative}(a) and (b). 
The $1^-_1$ state is dominated by the $K1^-$ component, 
which contributes to the $K^\pi=1^-$ band structure, whereas the $1^-_2$ state 
contains larger $K0^-$ component than the $K1^-$ component.
Note that these two states have significant $K$-mixing and shape fluctuation along $\beta$.
In high-lying negative-parity spectra, 
$K^\pi=1^-$ and $K^\pi=0^-$ bands are formed from the 
$1^-_6$ and $1^-_9$ states, respectively. These bands are 
formed by largely deformed bases with 
developed cluster structures, and they can be understood as cluster bands, which we label as $K^\pi=1^-_\textrm{cl}$ and 
$K^\pi=0^-_\textrm{cl}$ bands, respectively. 
The $K^\pi=0^-_\textrm{cl}$ band has a remarkable cluster structure 
of the $K0^-$ bases in $\beta >0.8$ region in particular. The dominant component of this 
state is the $K0^-(0.84)$ base (Fig.~\ref{fig:matter_density_negative}(d)), which has a developed $\Cs+\alpha$ structure similar to 
the $K^\pi=0^+_{\textrm{cl},2}$ cluster band, and therefore the $K^\pi=0^+_{\textrm{cl},2}$ and 
 $K^\pi=0^-_{\textrm{cl}}$ are 
regarded as the parity partner states of the $\Cs+\alpha$ clustering.
However, 
the $K^\pi=1^-_\textrm{cl}$ band is dominated by the $K1^-(0.64)$ base (Fig.~\ref{fig:matter_density_negative}(c)) 
with a weaker cluster structure than the $K^\pi=0^-_{\textrm{cl}}$  band.


Although the experimental information for negative-parity states is 
not enough to allocate band structures,
we tentatively allocate present $1^-_1$ and $1^-_2$ states to the experimental 
$1^-_1(5.36~\textrm{MeV})$ and $1^-_2(6.84~\textrm{MeV})$ states.
The  $E3$ transition 
from the $3^-\ (5.62\ \textrm{MeV})$ state to the $0^+_1$ state
was observed to have a significant strength of 
$B(E3)=170 \pm 14$\ $e^2\textrm{fm}^6$~\cite{Nakatsuka:2017dhs}.
We obtain $B(E3;3^-_1\to 0^+_1)=87.5$ 
$e^2\textrm{fm}^6$ between the $K^\pi=1^-_1$ and ground bands in this result. This value is of the 
same order as the experimental data and supports 
our conclusion that our $K^\pi=1^-_1$ band corresponds to the experimental $1^-_1~(5.36~\textrm{MeV})$ and 
$3^-\ (5.62\ \textrm{MeV})$ states.
For  dipole transition strengths from the ground to low-lying $1^-$ states, 
we will show the result 
in Sect.~\ref{sec:dipole} for discussions of dipole transition properties.

\subsection{Single-particle states in deformed states}

\begin{figure}
\begin{center}
\includegraphics[width=\hsize]{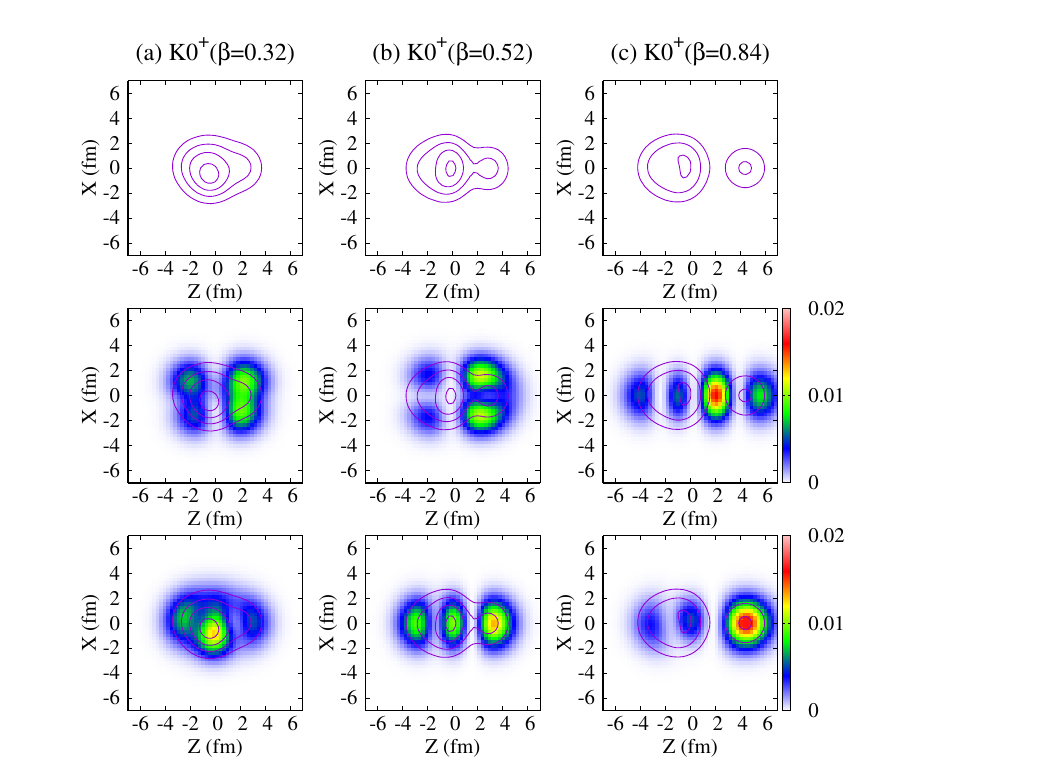} 
\end{center}
\caption{
Density distribution of protons and those of single-particle orbits in the (a)~$K0^+(0.32)$, (b)~$K0^+(0.52)$,
and (c)~$K0^+(0.84)$ bases, which correspond to the $\Kp=0^+_1$, $\Kp=0^+_\textrm{cl,1}$, 
and $\Kp=0^+_\textrm{cl,2}$ bands, respectively. 
The upper panels show the proton density distributions by the contour lines.
In the middle and lower panels,  
the density of the highest neutron and proton orbits are shown with color maps, respectively, 
with the total proton density (contour lines).  
The matter densities of these bases are shown in 
Fig.~\ref{fig:matter_density_positive}.
}
\label{fig:SP_density_positive}
\end{figure}

\begin{figure}
\begin{center}
\includegraphics[width=\hsize]{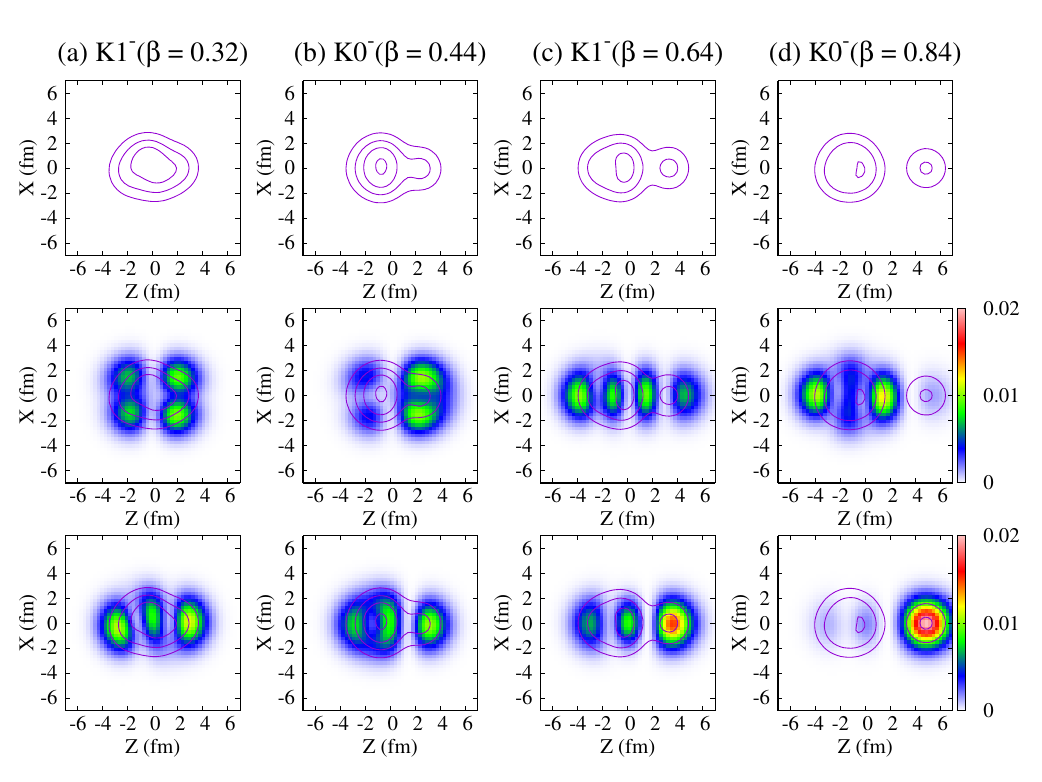} 
\end{center}
\caption{
The same as Fig.~\ref{fig:SP_density_positive}, but the results for the 
(a)~$K1^-(0.32)$,  (b)~$K0^-(0.44)$,
(c)~$K1^-(0.64)$, and (d)~$K0^-(0.84)$ bases, which correspond 
to the $\Kp=1^-_1$ band, the $1^-_2$ state, $\Kp=1^-_\textrm{cl}$, 
and $\Kp=0^-_\textrm{cl}$ bands, respectively. 
The matter densities of these bases are shown in 
Fig.~\ref{fig:matter_density_negative}.
}
\label{fig:SP_density_negative}
\end{figure}

To investigate single-particle configurations in a mean field picture,
we analyze single-particle orbits in the dominant components of the band-head $0^+$ and $1^-$ states and the $1^-_2$ state.
For each base, the wave function  
is expressed by a single Slater determinant, for which 
the nonorthogonal set of Gaussian single-particle wave functions can be transformed into 
an orthogonal set of single-particle orbits
in a mean-field as done in  Refs.~\cite{Dote:1997zz,Kanada-Enyo:1999bsw}.
Figures~\ref{fig:SP_density_positive} and \ref{fig:SP_density_negative} 
show single-particle orbits in the dominant bases of the positive- and negative-parity states,
respectively. 
For each basis, single-particle densities (color maps) of the highest neutron and proton orbits are 
shown together with the total proton density (contour lines).


Figures~\ref{fig:SP_density_positive}(a), (b), and (c) show results of 
the $K0^+(0.32)$, $K0^+(0.52)$, and $K0^+(0.84)$ for the $\Kp=0^+_1$, $\Kp=0^+_{\textrm{cl},1}$, 
and  $\Kp=0^+_{\textrm{cl},2}$ bands, respectively. 
The $K0^+(0.32)$ base for the $\Kp=0^+_1$ band is described by four neutrons in $sd$-orbits
around a weakly deformed core of the $\Os$ ground state, and it roughly corresponds to a $0\hbar\omega$
shell-model configuration.
The $K0^+(0.52)$ base for the $\Kp=0^+_{\textrm{cl},1}$ band has the character of two-proton excitation 
$p^{-2}_\pi (sd)^2_\pi$ of a $2\hbar\omega$ configuration in terms of the mean-field picture. 
In the cluster picture, the proton structure of this band has a parity asymmetric 6+2 structure and analogous to 
the proton part of the $\Os(0^+_2)$ state having a $\Ct+\alpha$ cluster structure.
The $K0^+(0.84)$ base for the $\Kp=0^+_{\textrm{cl},2}$ band has the 
developed $\Cf+\alpha$-cluster core with two 
neutrons in an elongated negative-parity orbit. This neutron orbit has three nodes along the $Z$ axis
and corresponds to a molecular called the $\sigma$-orbit. We label this negative-parity $\sigma$ orbit as
 $\sigma_{fp}$ in the association of a $fp$-orbit.
The $K0^+(0.84)$ base is associated with 
$4\hbar\omega$ configuration with  two-proton and neutron excitation in the mean-field picture. 
Note that,  after the GCM calculation, 
the final wave function of the $\Kp=0^+_{\textrm{cl},2}$ band contains not only the  $K0^+(0.84)$ component but 
also significant mixing of $K0^+(\beta > 0.84)$ bases with
the last two neutrons not in the molecular $\sigma_{fp}$-orbit but localized around the $\Cf$ cluster 
forming a dinuclear structure of $\Cs+\alpha$ cluster.
It means that 
the $\Kp=0^+_{\textrm{cl},2}$ band is a mixture of two types of clustering. One is 
the molecular orbital structure of 
the $\Cf+\alpha$ cluster core with two neutrons in the $\sigma_{fp}$-orbit and 
the other is the dinuclear $\Cs+\alpha$ structure.

Figures \ref{fig:SP_density_negative}~(a), (b), (c), and (d) present the results for
the $K1^-$($0.32$), $K0^-$($0.44$),  $K1^-$($0.64$), and $K0^-$($0.84$)
bases, which correspond to the $1^-_1$ and $1^-_2$ states, and the $\Kp=1^-_{\textrm{cl}}$ and 
$\Kp=0^-_{\textrm{cl}}$ bands, respectively. 
The $K1^-$($0.32$) base for the $1^-_1$ state can be understood as one proton excitation from the 
$p$ shell and associated with the $(1,0,0)^{-1}(0,0,2)^{1}$ (or $(0,1,0)^{-1}(0,0,2)^{1}$) configuration 
in terms of harmonic oscillator orbits $(n_x,n_y,n_z)$.
Furthermore, the $K1^-$(0.44) base for the $1^-_2$ state has one proton excitation 
as a leading component but cannot be interpreted by a simple $1p1h$ configuration. Instead, the proton excitation induces 
the parity asymmetric collective excitation in the proton and neutron parts
as can be seen in the asymmetry of the highest neutron orbit and that of the 
proton density in Fig.~\ref{fig:SP_density_negative}(b).
The $K1^-(0.64)$ base for the  $\Kp=1^-_{\textrm{cl}}$ band corresponds 
to a $3\hbar\omega$ excitation with 
one neutron in the $\sigma_{fp}$-orbit around the 
developed cluster core having two-proton excitation.
The $K0^-(0.84)$ base for the $\Kp=0^-_{\textrm{cl}}$ band has the dinuclear structure of developed $\Cs+\alpha$ clustering. 

Let us compare the intrinsic configurations the positive- and negative-parity cluster bands; 
the $\Kp=0^+_{\textrm{cl},1}$, $\Kp=0^+_{\textrm{cl},2}$, $\Kp=1^-_{\textrm{cl}}$, and $\Kp=0^-_{\textrm{cl}}$ 
bands. In these four cluster bands, the proton density has asymmetric shapes due to the 6+2 structure  
and shows ${}^{n}\textrm{C}+\alpha$ clustering. In terms of  the neutron configuration, 
the $\Kp=0^+_{\textrm{cl},1}$, $\Kp=1^-_{\textrm{cl}}$, and $\Kp=0^+_{\textrm{cl},2}$ bands have 
zero, one, and two neutrons in the $\sigma_{fp}$-orbit around the cluster core, respectively.
 As the number of $\sigma_{fp}$-orbit neutrons increases from zero to two, the 
cluster structure develops. 
It is interesting that the $\Kp=0^+_{\textrm{cl},2}$ band also contains significant mixing of the $\Cs+\alpha$ component, 
which is the dominant component of the $\Kp=0^-_{\textrm{cl}}$ band. 
Therefore, an alternative interpretation is that 
the $\Kp=0^+_{\textrm{cl},2}$ and $\Kp=0^-_{\textrm{cl}}$ bands form 
parity doublet partners of the $\Cs+\alpha$ structure.

\section{Discussions}\label{sec:discussion}
\subsection{Properties of dipole excitations} \label{sec:dipole}
\subsubsection{Dipole transition strengths}

\begin{figure}
\begin{center}
\includegraphics[width=\hsize]{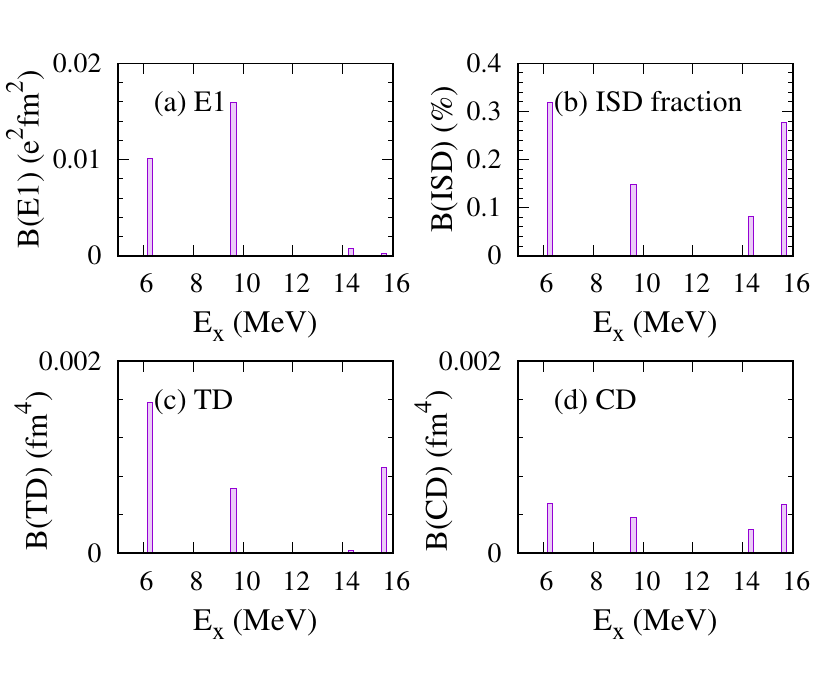} 
\end{center}
\caption{Dipole transition strengths for the (a)~$E1$, (b)~ISD, (c)~TD, and (d)~CD operators from the 
$0_1^+$ state.
For the ISD operator,  the energy weighted strengths are 
plotted in ratio to the EWSR defined by Ref.~\cite{Harakeh:1981zz}.
}
\label{fig:LED_strength}
\end{figure}


\begin{table}[htbp]
\begin{center}
\caption{\label{tab:dipole}
Calculated and experimental values of the excitation energies of the $1_1^-$ and $1_2^-$ states 
and the $E1$ and ISD transition strengths from the $0^+_1$ state.
For the ISD transitions, 
the energy weighted strength ratios ($f_\textrm{ISD}$) to the EWSR are listed.
The experimental data are taken from Refs.~\cite{Tryggestad:2002mxt,Tryggestad:2003gz,Nakatsuka:2017dhs}.
}
\begin{tabular}{cccc} \hline
&\multicolumn{3}{c}{Calculation}\\
&$E_x$\ (MeV)&$B(E1)$\ ($e^2\textrm{fm}^2$)&$f_{\textrm{ISD}}$\ ($\%$)\\
$1_1^-$&6.25&$1.11\times 10^{-2}$&0.31\\
$1_2^-$&9.59&$1.56\times 10^{-2}$&0.15\\
\\
&\multicolumn{3}{c}{Experiment}\\
&$E_x$\ (MeV)&$B(E1)$\ ($e^2\textrm{fm}^2$)&$f_{\textrm{ISD}}$\ ($\%$)\\
$1_1^-$&5.36(5)&$3.57(20)\times 10^{-2}$&2.70(32)\\
$1_2^-$&6.84(7)&$3.79(26)\times 10^{-2}$&0.67(12)\\
\hline
\end{tabular}
\end{center}
\end{table} 

The dipole transition strength function from the $0^+_1$ state 
is calculated using the
$0^+_1$ and $1^-_k$ states obtained with the GCM calculation. Figure~\ref{fig:LED_strength}~(a) shows the 
$E1$ strengths.
The energy-weighted ISD strengths are 
plotted in ratio to the EWSR as shown Fig.~\ref{fig:LED_strength}~(b).
The transition strengths for the CD and TD operators are shown in Figs.~\ref{fig:LED_strength}~(c) and (d), 
respectively. Significant $E1$ and TD transition strengths are obtained for the 
two LED states, $1^-_1$ and $1^-_2$ states. The $1^-_1$ state has a remarkable TD and 
significant $E1$ strengths, whereas the $1_2^-$ state has remarkable $E1$ strength.
Compared with the TD strengths, the CD transitions to the two LED states are rather weak as 
0.3\% (0.15\%) of the EWSR  for the $1^-_1$  ($1^-_2$) states.
In Table~\ref{tab:dipole}, we compare the present results of the  $E1$ and ISD transition strengths 
to the $1^-_1$ and $1^-_2$ states with the experimental data 
of the $1^-_1$(5.36 MeV) and $1^-_2$(6.84 MeV) states.
This result  qualitatively 
describes 
the significant $E1$ strengths 
measured for the $1^-_1$(5.36 MeV) and $1^-_2$(6.84 MeV) states, 
though the quantitative agreement with the data is not satisfactory.
For the ISD strengths,  this calculation fails to obtain significant ISD strengths as
large as the observed ISD strength to the $1^-_1$ state reported recently~\cite{Nakatsuka:2017dhs}.
Our result for weak ISD transitions to LED states agrees to 
a mean-field calculation~\cite{Inakura:2018ccl}.

\subsubsection{Transition current and strength densities for LED in $\Ot$}

We calculate the transition current and strength densities
in the intrinsic frame
using the dominant bases
to discuss the properties of the low-energy dipole excitations $0^+_1\to 1^-_{1,2}$.
The definitions for the transition current and strength densities are 
given in appendix \ref{app:1}.
For the intrinsic states of the $0^+_1$, $1^-_1$, and $1^-_2$ states, 
we choose the $K0^+(0.32)$, $K1^-(0.32)$, and 
$K0^-(0.44)$ bases, respectively, to describe the leading properties of 
each state, 
and calculate the transition current densities of the $K0^+(0.32)\to K1^-(0.32)$
and $K0^+(0.32)\to K0^-(0.44)$ transitions.
In the calculation, normalized $K$ eigenstates projected from the wave functions
$\Phi^\pi_{K}(\beta)$ are used 
as explained in appendix~\ref{app:1}.
Note that, the $1^-_1$, and $1^-_2$ states significantly 
contain the $K$-mixing and shape fluctuation along $\beta$, which 
contributes to the final GCM results of the 
 $1^-_1$, and $1^-_2$ states, 
but such higher order effects are omitted for simplicity in the this analysis in the 
intrinsic frame.

\begin{figure}
\begin{center}
\includegraphics[width=\hsize]{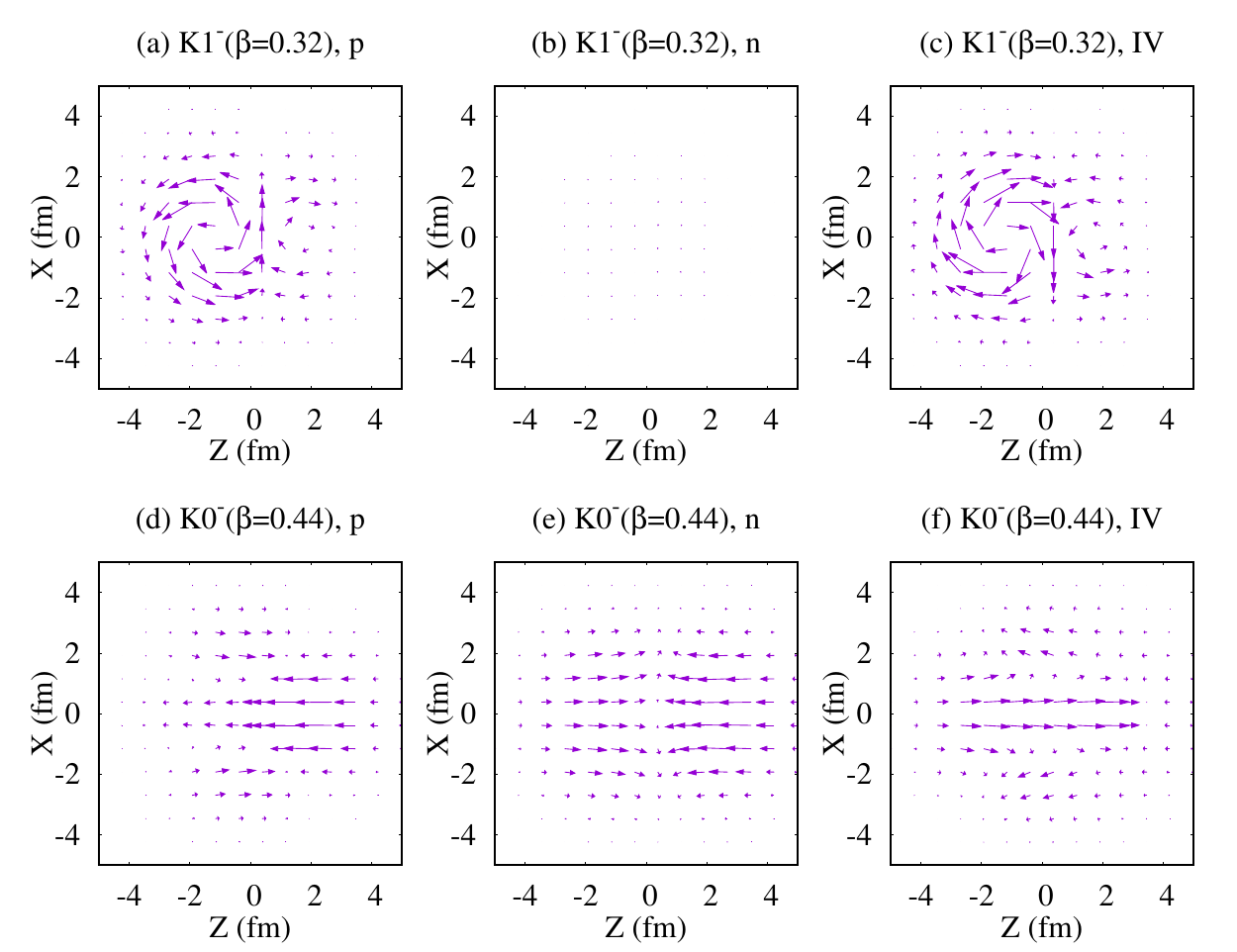} 
\end{center}
\caption{
(upper) Transition current densities $\delta\vector{j}^K(\vector{r})$
from the $K0^+(0.32)$ base to the 
$K1^-(0.32)$ corresponding to 
the $0^+_1\to 1^-_1$ transition and (lower) those to the $K0^-(0.44)$ base 
for $0^+_1\to 1^-_2$.
The vector plots of the densities in the $Z$-$X$ plane at $Y=0$ are shown.
The proton and neutron currents are shown in the left and middle, respectively, 
and the isovector currents are shown in the right.
The vector plots are multiplied by 30 in (a)-(c), 
by 50 in (d) and (e), and by 100 in panel (f).
}
\label{fig:transition_current}
\end{figure}

\begin{figure}[t]
\begin{center}
\includegraphics[width=\hsize]{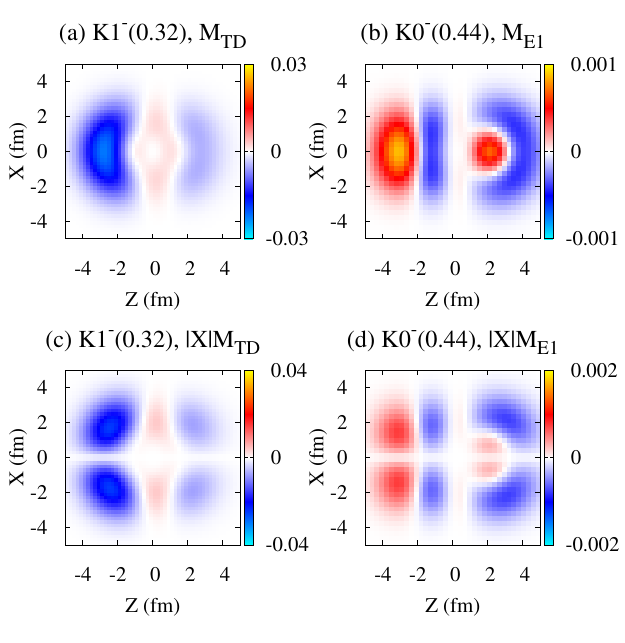} 
\end{center}
\caption{
(Left) TD strength densities ${\cal M}^K_{\textrm{TD}}$ for 
$K0^+(0.32)\to K1^-(0.32)$ and (Right) $E1$ strength densities ${\cal M}^K_{E1}$
for $K0^+(0.32)\to K0^-(0.44)$. 
The former and the latter correspond to the $0^+_1\to 1^-_1$ and 
$0^+_1\to 1^-_2$ transitions, respectively.
The strength densities ${\cal M}^K(X,Y,Z)$ and $|X|$-weighted values 
$|X|{\cal M}^K(X,Y,Z)$  on the $Z$-$X$ plane 
at $Y=0$ are shown in upper and lower panels, respectively. 
}
\label{fig:strength_density}
\end{figure}

The calculated transition current densities are shown in Fig.~\ref{fig:transition_current}.
Vector plots in the left, middle, and right panels show 
the proton and neutron parts and the isovector component of the transition current densities,
respectively. 
The strength densities of the TD and $E1$ operators are shown in Fig.~\ref{fig:strength_density}.
The vortical flow of the proton current density is induced by 
the 1 proton excitation $(1,0,0)^{-1}(0,0,2)^1$ 
in the $K0^+(0.32)\to K1^-(0.32)$ transition, which corresponds to the $1^-_1$ excitation 
as shown in the transition current density
in Fig.~\ref{fig:transition_current}(a). 
This vortical proton current contributes to the remarkable TD strength density 
as shown in Fig.~\ref{fig:strength_density}(a) and describes the TD nature of  the $1^-_1$ excitation.
However, the $K0^+(0.32)\to K0^-(0.44)$ transition for the $1^-_2$ excitation 
show a translational flow along the deformed ($Z$) axis 
rather than a vortical flow
(see Fig.~\ref{fig:transition_current}(b)).
The neutron part of the translational flow, in particular, is widely distributed across a wide $X$ range. 
The surface neutron flow in the region of $|X|=2$--4~fm and $Z\sim 2$~fm 
is produced by valence neutron oscillation
in the parity asymmetric orbit (Fig.~\ref{fig:SP_density_negative}(b)) around the prolate core, 
which is induced by the proton excitation. 
This neutron surface flow, as shown in Fig.~\ref{fig:strength_density}(d), 
gives the dominant contribution to the $E1$ strength
of the $K0^+(0.32)\to K0^-(0.44)$ transition and is a major source of the strong $E1$ transition to 
the $1^-_2$ state.
In the internal region of the prolately deformed core, the proton and neutron flows  
cancel each other, but give some contribution to the $E1$ strength 
because of the recoil effect. This result indicates that
the parity asymmetry of the cluster core and that of the valence neutron orbit, 
which are induced by the two-proton excitation, play an important role in the enhanced 
$E1$ strength of the $K0^-(0.44)$ base. 

In the this analysis of the $K1^-(0.32)$ and $K0^-(0.44)$ bases, 
a clear difference is found in the transition properties between the two LED modes;
the TD nature in the $K1^-(0.32)$ base and the $E1$ character in the $K0^-(0.44)$ base.
These two LED modes, the TD and $E1$ modes appear separately 
as vortical and translational excitations of nuclear current 
in the $\Kp=1^-$ and 
$\Kp=0^-$ components of the deformed states, respectively. However 
they couple with each other 
in the $1^-_1$ and $1^-_2$ states after the superposition of the GCM calculation
via significant $K$-mixing and shape fluctuation as mentioned previously. 
Therefore, the TD strength of the $K1^-(0.32)$ base is fragmented into
the $1^-_1$ and $1^-_2$ states, and the $E1$ strength of the $K0^-(0.44)$ base is split into 
the two states. 
Nevertheless, 
since the $1^-_1$ state retains the dominant TD nature, 
it has a relatively large TD strength and constructs the $K^\pi=1^-$ band structure.

\subsection{Systematic analysis of LED excitations in O isotopes}

\begin{figure*}[t]
\begin{center}
\includegraphics[width=\hsize]{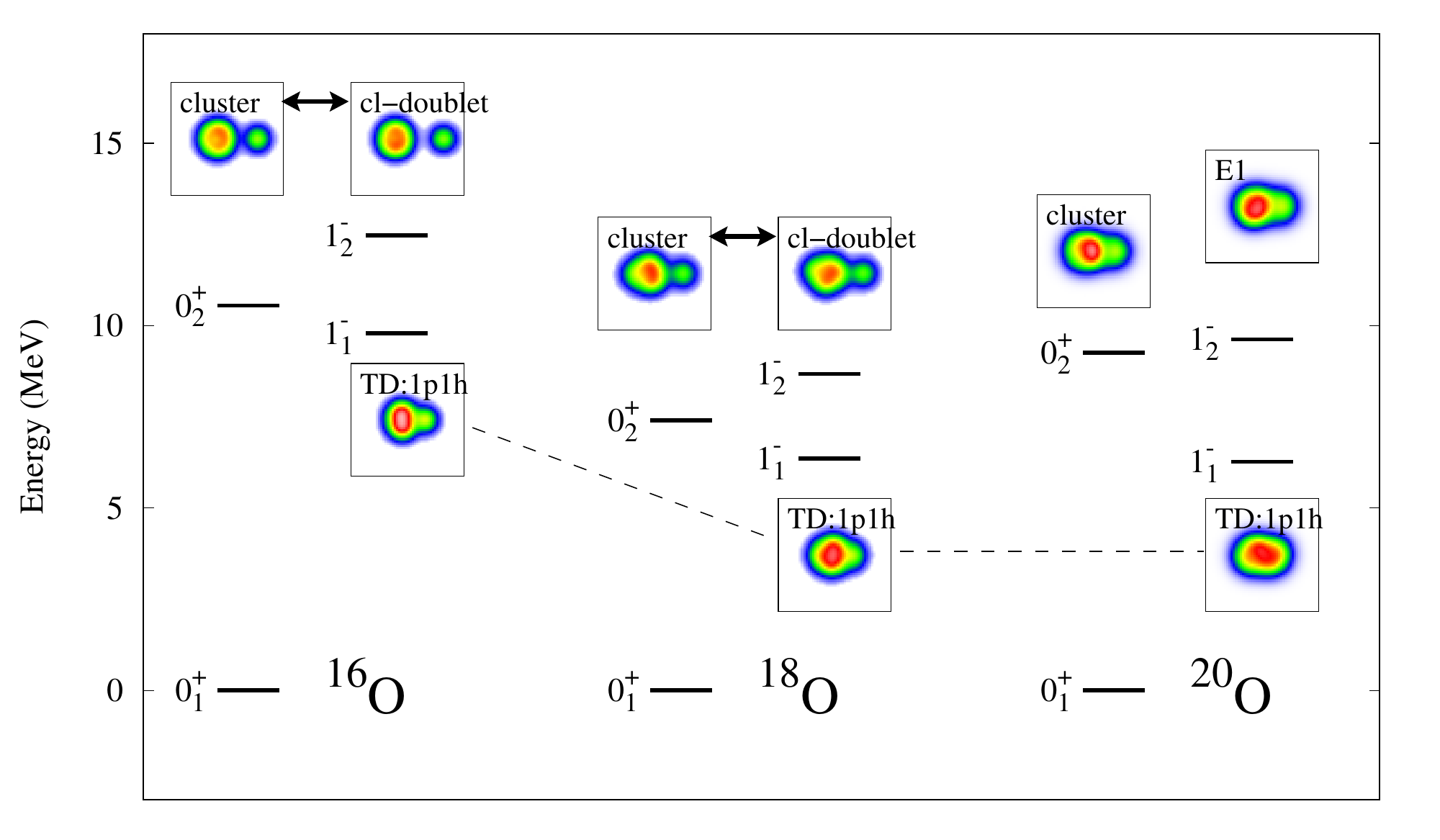} 
\end{center}
\caption{Energy spectra of the $0^+_{1,2}$ and $1^-_{1,2}$ states in $\Os$, $\Oe$, and $\Ot$
calculated with K-VAP and GCM of $\beta$-constraint AMD. For excited states, 
intrinsic matter densities of the dominant bases
are also shown with labels ``TD:1p1h'', ``cluster'', ``cluster-doublet'', and ``E1'', which indicate 
the TD mode with $1p$-$1h$ configuration, $\Kp=0^+$ cluster state, its parity doublet $\Kp=0^-$ state, and the $E1$ mode, respectively. 
The color map plotting of densities is the same as 
Fig.~\ref{fig:matter_density_positive}.} 
\label{fig:oisotopes-spe}
\end{figure*}

\begin{figure}[t]
\begin{center}
\includegraphics[width=\hsize]{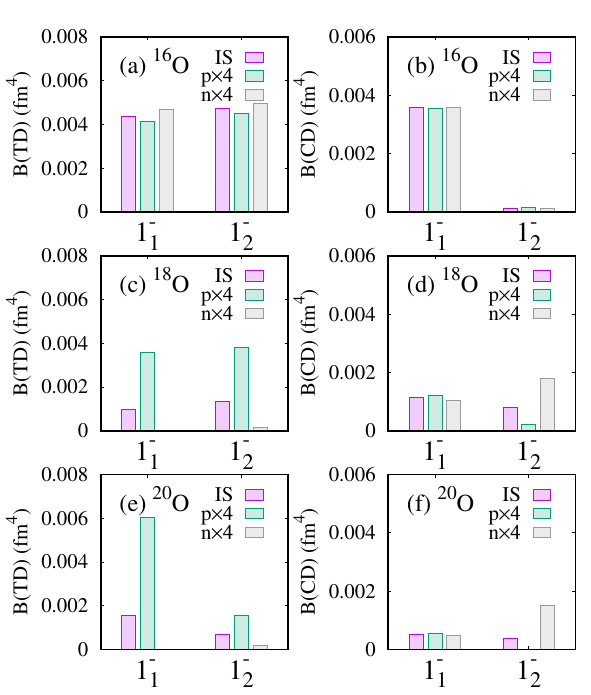} 
\end{center}
\caption{
The IS, proton, and neutron components of 
the TD and CD strengths for the $1^-_1$ and $1^-_2$ states of $\Os$, $\Oe$, and $\Ot$ 
calculated with K-VAP and GCM of $\beta$-constraint AMD.
The TD strengths of (a) $\Os$, (c) $\Oe$, and (e) $\Ot$ are shown in the left, 
and the CD strengths of (b)~$\Os$, (d)~$\Oe$, and (f)~$\Ot$ are shown in the right. 
Proton and neutron components are multiplied by a factor of four
to compare the IS component. 
The results for $\Os$ and $\Oe$ are taken from Refs.~\cite{Shikata:2020lgo,Shikata:2020iez}.
}
\label{fig:oisotopes-strength}
\end{figure}

\begin{figure}[t]
\begin{center}
\includegraphics[width=\hsize]{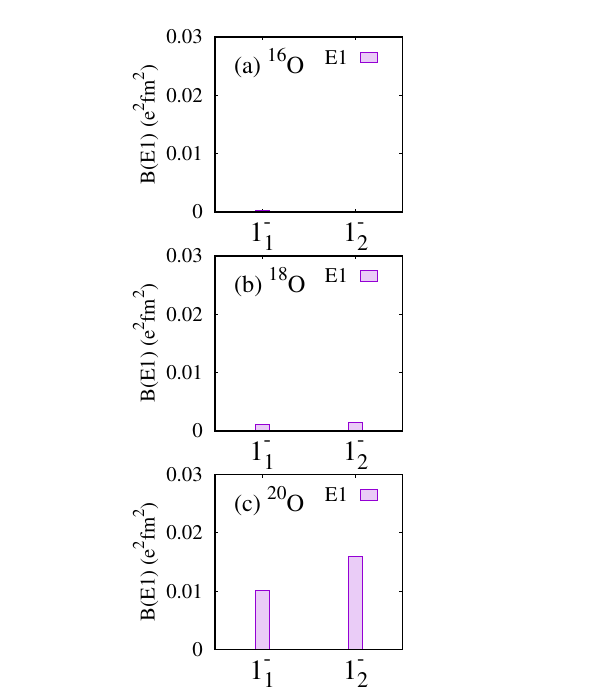} 
\end{center}
\caption{$E1$ strengths for the $1^-_1$ and $1^-_2$ states of (a)$\Os$, (b)$\Oe$, and (c)$\Ot$ calculated with 
K-VAP and GCM of $\beta$-constraint AMD. 
The results for $\Oe$ are taken from Ref.~\cite{Shikata:2020iez}.
}
\label{fig:oisotopes-strength-E1}
\end{figure}

To clarify the roles of valence neutrons in the LED excitations in $\Ot$,
we discuss systematics of dipole excitation properties in O isotopes by comparing 
the present results with previous results obtained using the same framework 
for $\Os$ and $\Oe$.
Figure \ref{fig:oisotopes-spe} shows 
the theoretical energy spectra of the $0^+_{1,2}$ and $1^-_{1,2}$ states
in $\Os$, $\Oe$, and $\Ot$. The intrinsic matter densities of the dominant bases in the excited states are
also shown in the figure.
In each of $\Os$, $\Oe$, and $\Ot$,
two $1^-$ states are obtained in the low-energy region. 

These LED states have 
significant isoscalar dipole strengths of the TD and/or CD operators. 
 Figure ~\ref{fig:oisotopes-strength} shows 
the isoscalar, proton, and neutron components of 
the TD and CD strengths for the $1^-_1$ and $1^-_2$ states of the O isotopes.
According to the previous analysis,
we identified the $\Os(1^-_1)$ and $\Oe(1^-_1)$ states as TD mode, which is characterized by 
the vortical flow of the transition current densities. These LED states in $\Os$ and $\Oe$ correspond to
the present TD mode of the $\Ot(1^-_1)$ state.
The TD mode is described by the $\Kp=1^-$ component of the 
$1p$-$1h$ excitation of deformed states in all three cases.
The isoscalar components of the TD strengths of the $1^-_1$ and $1^-_2$ states are largest in $\Os$ because of the 
coherent (isoscalar) contribution from the proton and neutron parts, but relatively small 
in $\Oe$ and $\Ot$ because of the lack of contribution from the neutron part.

Figure~\ref{fig:oisotopes-strength-E1} shows 
the $E1$ strengths for the $1^-_1$ and $1^-_2$ states of the O isotopes.
The low-energy $E1$ mode is obtained only in the $\Ot$ as the $\Ot(1^-_2)$ state, 
which is produced by the previously described surface neutron oscillation on the prolate deformation induced by 
proton excitation.
The $\Os(1^-_2)$ and $\Oe(1^-_2)$ states are not $E1$ modes but have a distinct character, 
that is, the asymmetric cluster structure that forms 
parity partners with the $\Os(0^+_2)$ and $\Oe(0^+_2)$ states, respectively.
Note that the $\Ot(0^+_2)$ state has a cluster structure but its parity doublet partner $1^-$ state is not obtained.
In the structure change from the $\Os(0^+_2)$ state along the isotope chain,
the clustering is weakened in the $\Oe(0^+_2)$ state and further suppressed in the $\Ot(0^+_2)$ state
by excess neutrons and no longer constructs the parity doublet $1^-$ state of the $\Ot(0^+_2)$ state.

Finally, we comment on the CD strengths in the LED states of O isotopes. 
As shown in Fig.~\ref{fig:oisotopes-strength}(a), the strong CD transition was obtained 
in the TD mode of $\Os$, which is consistent with the
experimental observation of the ISD strength of the $\Os(1^-_1)$. 
However, the present calculation does not degenerate such a strong CD strength in the TD mode 
of $\Ot$, and fails to describe the observed ISD strength of the $\Ot(1^-_1)$ state. 
According to the previous analysis in Ref.~\cite{Shikata:2020lgo}, 
the origin of the strong CD transition in the $\Os(1^-_1)$ state is 
significant $K$-mixing of the TD mode and coupling with other deformed bases 
via the $\beta$ fluctuation. The contribution of the CD strengths contained in the 
$K0^-$ component of the normal deformation is essential. 
However, in the present result of $\Ot$, the low-lying $E1$ 
appears in the $K0^-$ component of the normal deformation, 
which contributes only weakly to the CD strength. 
In the present calculation of GCM along the $\beta$ deformation, 
only the lowest base at each $\beta$ is obtained by the energy optimization, and thus
energetically higher bases containing the CD strength may be missing. 
To overcome this problem, it is necessary to extend the present framework to properly 
include important bases for the low-lying CD strengths.

\section{Summary}\label{sec:summary}

K-VAP and GCM of $\beta$-constraint AMD were used to investigate LED excitations in $\Ot$.
Two LED states, the $1^-_1$ and $1^-_2$ states were obtained. 
The $1^-_1$ state is a weakly deformed state with 
one-proton excitation, whereas the $1^-_2$ state has a normal deformation
with the parity asymmetric structure. 

In a detailed analysis of the dipole transition properties of these LED states, 
the $1^-_1$ state is considered the TD mode, while the $1^-_2$ state is associated with a low-energy $E1$ mode. 
The TD strength in the former mode is produced 
by vortical nuclear current, whereas the $E1$ strength in the latter mode  is 
contributed by surface neutron current on the prolate deformation induced by proton excitation. 
These two modes, the TD (vortical) and $E1$ modes, appear  separately as the $\Kp=1^-$ and $\Kp=0^-$ components of 
the deformed states, but they couple with each other in the $1_1^-$ and $1_2^-$ states of $\Ot$ via
the $K$-mixing and shape fluctuation along $\beta$. 
Therefore, the TD and $E1$ strengths are fragmented into both $1^-$ states.

In comparison with the experimental data of the $E1$ and ISD transition strengths to the $1^-_1$(5.36~MeV)
and the $1^-_2$(6.84~MeV) states, 
the present calculation qualitatively described the experimental $E1$ strengths for the $1^-_1$ and $1^-_2$
states, but much underestimated the significant ISD transition strengths observed for the  $1^-_1$ state by 
one order. 

To clarify the roles of valence neutrons in LED excitations in $\Ot$, 
systematics of the LED excitations in $\Os$, $\Oe$, and $\Ot$ were discussed in comparison 
for the present $\Ot$ result with the previous $\Os$ and $\Oe$ results 
obtained using the same framework. 
The TD mode was obtained as the lowest $1^-_1$ state in $\Os$, $\Oe$, and $\Ot$. However,
the low-energy $E1$ mode was found only in the $\Ot(1^-_2)$ state but not in the $\Os$ and $\Oe$ systems.
Instead, the previous results indicated that the $\Os(1^-_2)$ and $\Oe(1^-_2)$ states differ from the $\Ot(1^-_2)$ state
and are parity doublet partners in the $\Kp=0^-$ cluster band 
with the $0^+_2$ states in the $\Kp=0^+$ bands.

\section*{Acknowledgment}
The computational calculations of this work were performed  using the
supercomputer at the Yukawa Institute for theoretical physics, Kyoto University. 
This work was supported by JSPS KAKENHI Grant Nos. 18J20926, 18K03617, and 18H05407.

\appendix

\section{Densities of intrinsic system in the body-fixed frame}\label{app:1}

Isoscalar and isovector components of the density and current density operators are defined as,  
\begin{eqnarray}
&&\rho(\vector{r}) = \sum_{k=1}^A \delta(\vector{r}-\vector{r}_k),\\
&&\rho_{\textrm{IV}} = 
\sum_{k=1}^A \frac{e^\textrm{eff}}{e}
 \delta(\vector{r}-\vector{r}_k) ,\\
&&\vector{j}_{\textrm{nucl}}(\vector{r})=\frac{-i\hbar}{2m} \nonumber\\
&&\quad \times \sum_{k=1}^A\{ \vector{\nabla}_k\delta(\vector{r}-\vector{r}_k) + \delta(\vector{r}-\vector{r}_k)\vector{\nabla}_k \},\\
&&\vector{j}_{\textrm{nucl},\textrm{IV}}(\vector{r}) =\frac{-i\hbar}{2m} \nonumber\\ 
&&\quad \times \sum_{k=1}^A \frac{e^\textrm{eff}}{e}\{ \vector{\nabla}_k\delta(\vector{r}-\vector{r}_k) + \delta(\vector{r}-\vector{r}_k)\vector{\nabla}_k \},
\end{eqnarray}
where the factor $e^\textrm{eff}/e$ is $N/A$ for protons and $-Z/A$ for neutrons.
The diagonal densities for $|k\rangle=|\Phi^\pi_K(\beta)\rangle$ are expressed as, 
\begin{eqnarray}
\rho(\vector{r}) &\equiv & \langle k| \hat \rho(\vector{r}) |k\rangle ,\\
\rho_{\textrm{IV}}(\vector{r}) &\equiv & \langle k| \hat\rho_{\textrm{IV}}(\vector{r})|k \rangle.
\end{eqnarray}
The transition densities and transition current densities for initial $|i\rangle$ and final $|f\rangle$ states 
are given as,
\begin{eqnarray}
\delta\rho(\vector{r}) &\equiv & \langle f|{\hat\rho}(\vector{r})|i\rangle ,\\
\delta\rho_{\textrm{IV}}(\vector{r}) &\equiv & \langle f|{\hat\rho}_{\textrm{IV}}(\vector{r})|i\rangle ,\\
\delta\vector{j}(\vector{r}) &\equiv & \langle f|\hat{\vector{j}}_{\textrm{nucl}}(\vector{r})|i\rangle ,\\
\delta\vector{j}_{\textrm{IV}}(\vector{r}) &\equiv & \langle f|\hat{\vector{j}}_{\textrm{nucl,IV}}(\vector{r})|i\rangle.
\end{eqnarray}

In the present calculation with K-VAP of $\beta$-constraint AMD, 
each intrinsic wave function $\Phi^\pi_K(\beta)$ for a $K0^+(\beta)$, $K0^-(\beta)$, or 
$K1^-(\beta)$ base 
is expressed by a Slater determinant, and its intrinsic densities are 
given as the diagonal densities calculated for $|k\rangle=|\Phi^\pi_K(\beta)\rangle$. 
The transition densities and transition current densities 
from a $K0^+(\beta_0)$ base
to $K1^-(\beta_1)$ and $K0^-(\beta_2)$ bases are calculated in the intrinsic (body-fixed) frame 
for the $K$-projected bases,
\begin{align}
&|i\rangle=\hat P^{K=0}|\Phi^+_{K=0}(\beta_0)\rangle,\\
&|f\rangle=\frac{\hat P^{K=-1}-\hat P^{K=1}}{\sqrt{2}}|\Phi^-_{K=1}(\beta_1)\rangle \equiv 
|f^{K=1}\rangle,\\
&|f\rangle=\hat P^{K=0}|\Phi^-_{K=0}(\beta_2)\rangle\equiv 
|f^{K=0}\rangle,
\end{align}
where $|i\rangle$ and  $|f\rangle$ are normalized as $\langle i|i\rangle=\langle f|f\rangle=1$ by definition.
The local matrix elements ${\cal M}_{\textrm{TD,}E1}^K(\vector{r})$ of the TD and $E1$ operators
are calculated at $Y=0$ on the $Z$-$X$ plane in the intrinsic frame as, 
\begin{eqnarray}
&&\hana{M}_{\textrm{TD}}^{K=0}(X,0,Z)= \frac{1}{10c}\sqrt{\frac{3}{4\pi}}\nonumber \\&&\qquad\times
\left[ (2X^2+Z^2)\delta j^{K=0}_Z - ZX\delta j^{K=0}_X  \right],\label{eq:SD_TD_0} \\ 
&&\hana{M}_{\textrm{TD}}^{K=1}(X,0,Z)= \frac{1}{10c}\sqrt{\frac{3}{4\pi}}\nonumber \\&&\qquad\times
\left[ (X^2+2Z^2)\delta j^{K=1}_X - ZX\delta j^{K=1}_Z \right],\label{eq:SD_TD_1} \\
&&\hana{M}_{E1}^{K=0}(X,0,Z) = \sqrt{\frac{3}{4\pi}}Z\delta\rho^{K=0}_{\textrm{IV}},\label{eq:SD_E1_0} \\
&&\hana{M}_{E1}^{K=1}(X,0,Z) = \sqrt{\frac{3}{8\pi}}X\delta\rho^{K=1}_{\textrm{IV}},\label{eq:SD_E1_1}
\end{eqnarray}
where $\delta\rho^K_{\textrm{IV}} = \langle f^K|{\hat\rho}_{\textrm{IV}}(\vector{r})|i\rangle$
and $\delta\vector{j}^K = \langle f^K|\hat{\vector{j}}_{\textrm{nucl}}(\vector{r})|i\rangle$ at $\vector{r}=(X,0,Z)$.
Note that $\hana{M}_{\textrm{TD}}^{K}(\vector{r})$ and $\hana{M}_{E1}^{K}(\vector{r})$
correspond to the integrand of the TD and $E1$ transition matrix elements and are termed TD and $E1$ strength densities, respectively, in this paper.

\bibliographystyle{apsrev4-1}
\bibliography{reference_Dthesis} 

\end{document}